# Spectral filtering effect of diffraction gratings with a lens coupling to optical fibers

**SEONJONG RYU[1], JINPYO JEONG[2], MINTAE KANG[3], TAEMIN SON[3], ANDY CHONG[3,4*]**

[1]Department of Physics, Pohang University of Science and Technology(POSTECH), Pohang 37673, Republic of Korea
[2]Department of Electrical Engineering, Pusan National University, Geumjeong-Gu, Busan 46241, Republic of Korea
[3]Department of Physics, Pusan National University, Geumjeong-Gu, Busan 46241, Republic of Korea
[4]Institute for Future Earth, Pusan National University, Busan, 46241, Republic of Korea
*chong0422@pusan.ac.kr

**We present a theoretical study of a spectral filter, which consists of a diffraction grating, a coupling lens, and an optical fiber. As the diffracted beam is highly dispersed spatially, coupling into an optical fiber naturally creates a Gaussian spectral filtering effect. Using ray transfer matrices, we derive simple equations to calculate the spectral filter bandwidth and the group velocity dispersion. This study offers insights for designing fiber-based spectral filters, particularly for mode-locked fiber lasers.**

## 1. Introduction

Spectral filters that transmit selective wavelengths can be achieved by a variety of methods. Among them, spectral filters utilizing diffraction gratings can provide narrow filtering bandwidths across a broad operating range [1]. While a diffraction grating spatially disperses frequencies, partially blocking the beam induces a spectral filtering effect. This spectral filtering technique is widely used in spectrometers and monochromators by blocking the spatially dispersed beam with a narrow slit [2].

Instead of using slits, a lens coupling into an optical fiber also induces narrow spectral filtering. This optical fiber-based spectral filter has been demonstrated in optical communications [3] and has proven valuable for mode-locked fiber lasers. For example, all-normal-dispersion (ANDi) fiber lasers require spectral filters with a bandwidth (BW) of ~10 nm for stable mode-locking [4]. Self-similar fiber lasers require even narrower (~4 nm) BW filters which have been achieved by coupling of grating dispersed beams into optical fiber via a lens [5]. It is essential to accurately characterize the intracavity spectral filters, as they can significantly influence the laser performance [6]. Moreover, the group velocity dispersion (GVD) induced by the filter can further impact the laser performance. Although the filtering concept is straightforward, the calculations of BW and GVD become complex since the fiber coupling further influences the filtering effect.

The spectral BW and GVD of such filters can be calculated numerically via commercial ray tracing software. However, a simple analytic method for calculating the filter BW and GVD has not been fully explored. In this work, we propose analytic equations for determining the BW and GVD of such filters. The goal is to provide simple formulas that allow users to calculate the filter BW easily by inputting filter parameters. The filter BW is derived using transfer matrices (ABCD matrices) with initial errors introduced by the diffraction grating. GVD can be determined by calculating the frequency-dependent optical path lengths. The results are presented for commonly used filter parameters in mode-locked fiber lasers.

## 2. Spectral filter bandwidth (BW)

The schematic of the spectral filter is illustrated in Figure 1. After diffracted by a grating, a Gaussian beam propagates in a free space over a distance L. The beam is then focused by a lens into a single-mode -fiber at focal length $f$. As shown in the figure, each wavelength reaches the fiber tip with different translational and angular misalignments. These deviations from the ideal fiber coupling induce the spectral filtering effect. Since the lens-to-fiber setup forms a device known as a collimator, we will refer to this spectral filter as a grating - collimator spectral filter.

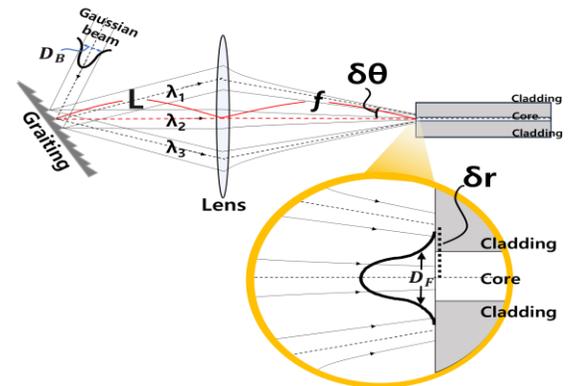

Figure 1. Spectral bandwidth calculation schematic

The ABCD matrix can be used to calculate the final location and angle of a ray after optical elements [7,8,9]. However, if the initial beam contains position and angle errors, such errors can be transmitted and even magnified in the output. This transmission of errors can be handled conveniently by manipulating a 3 x 3 matrix, known as the ABCDEF matrix which takes the form in Equation 1 [10].

$$M = \begin{bmatrix} A & B & E \\ C & D & F \\ 0 & 0 & 1 \end{bmatrix} \quad (1)$$

The ABCDEF matrix already contains errors, with the E and F components representing the translational and angular errors, respectively. The final output ray can be calculated conveniently using Equation 2. In other words, if errors are clearly defined, one can predict the output accurately as a function of those errors.

$$\begin{bmatrix} r_{out} \\ \theta_{out} \\ 1 \end{bmatrix} = \begin{bmatrix} A & B & \delta r \\ C & D & \delta \theta \\ 0 & 0 & 1 \end{bmatrix} \begin{bmatrix} r_{in} \\ \theta_{in} \\ 1 \end{bmatrix} \quad (2)$$

As shown in Figure 1, we need ABCDEF matrices for the free space propagation, the lens, and the diffraction grating. This is represented in Equations 3, 4, and 5, respectively. In Equation 3, $x$ represents the propagation distance, and $f$ is the focal length of the lens in Equation 4. $\Delta\theta$ denotes the difference in diffraction angle between the center wavelength $\lambda_c$ and an arbitrary wavelength $\lambda$ in Equation 5. It is important to note that the diffraction angle difference, calculated using the grating equation, can be introduced as the angular error as the function of wavelength $\lambda$ in the matrix.

$$M_{Freespace}(x) = \begin{bmatrix} 1 & x & 0 \\ 0 & 1 & 0 \\ 0 & 0 & 1 \end{bmatrix} \quad (3)$$

$$M_{Lens} = \begin{bmatrix} 1 & 0 & 0 \\ -\frac{1}{f} & 1 & 0 \\ 0 & 0 & 1 \end{bmatrix} \quad (4)$$

$$M_{Grating} = \begin{bmatrix} 1 & 0 & 0 \\ 0 & 1 & \Delta\theta(\lambda) \\ 0 & 0 & 1 \end{bmatrix} \quad (5)$$

From the grating equation $d(\sin\theta_i + \sin\theta) = m\lambda$, where d is the grating groove spacing, m is the diffraction order, $\theta_i$ is the incident angle, and $\theta$ is the diffraction angle, the diffraction angle difference $\Delta\theta$ can be derived as shown in Equation 6. In this equation, $\Delta\theta$ is Taylor expanded in terms of $\lambda - \lambda_c$. For this spectral filter case when $\lambda$ is close to $\lambda_c$ (i.e. ~10 nm spectral filter BW), the second order term is approximately ~$10^{-3}$ of the first order term and therefore, higher order terms are neglected.

$$\Delta\theta = \sin^{-1}\left(\frac{m\lambda - d\sin\theta_i}{d}\right) - \sin^{-1}\left(\frac{m\lambda_c - d\sin\theta_i}{d}\right)$$

$$\simeq \frac{m(\lambda - \lambda_c)}{d\sqrt{1 - \left(\frac{m\lambda_c - d\sin\theta_i}{d}\right)^2}} \quad (6)$$

Since the initial error is a function of wavelength, the final translational and angular errors at the fiber tip naturally become functions of wavelength $\lambda$. By multiplying a series of matrices, the wavelength dependency of the translational ($\delta r$) and angular error ($\delta\theta$) at the tip of the fiber can be determined by extracting the E and F components of the final matrix $M_{AtFiber}$, as shown in Equation 7.

$$M_{AtFiber} = M_{Freespace}(f) \cdot M_{Lens} \cdot M_{Freespace}(L) \cdot M_{Grating}$$

$$= \begin{pmatrix} 0 & f & \dfrac{\Lambda f(\lambda - \lambda_c)}{\sqrt{1 - \Lambda^2\left(\lambda_c - \dfrac{\sin(\theta_i)}{\Lambda}\right)^2}} \\ -\dfrac{1}{f} & 1 - \dfrac{L}{f} & -\dfrac{\Lambda(\lambda - \lambda_c)(L/f - 1)}{\sqrt{1 - \Lambda^2\left(\lambda_c - \dfrac{\sin(\theta_i)}{\Lambda}\right)^2}} \\ 0 & 0 & 1 \end{pmatrix} \quad (7)$$

Since the beam size influences the coupling efficiency and, consequently the filtering effect, the wavelength-dependent beam size at the tip of the fiber needs to be considered. ABCD matrix formalism for Gaussian beams can be applied to calculate the beam size according to Equation 8 [9].

$$q_{out} = \frac{Aq_{in} + B}{Cq_{in} + D} \qquad \frac{1}{q} = \frac{1}{R} - \frac{i\lambda_c}{\pi w^2} \quad (8)$$

In Equation 8, $q$ is the complex beam parameter, $R$ is the wavefront radius of curvature, $\lambda_c$ is the center wavelength, and $w$ is the beam radius. Since $q$ is not a strong function of wavelength for our filter case where $\lambda \sim \lambda_c$ (i.e. ~10 nm spectral filter BW), we used the center wavelength $\lambda_c$ to approximate $\lambda$. However, ABCD components extracted from $M_{AtFiber}$, contain the wavelength information necessary to determine the wavelength-dependent beam size. As the incoming Gaussian beam is focused to the beam radius of $w_B$ with translational misalignment $\delta r$ and angular misalignment $\delta\theta$, the single-mode-fiber coupling efficiency is given by Equation 9 and 10 [11].

$$\eta_0 = \frac{4}{\left(\dfrac{w_B}{w_F} + \dfrac{w_F}{w_B}\right)^2} \exp[\phi] \quad (9)$$

$$\phi = -\frac{2}{w_B^2 + w_F^2}\delta r^2 - \frac{2\pi^2 w_B^2 w_F^2}{\lambda_c^2(w_B^2 + w_F^2)}\delta\theta^2 \quad (10)$$

In Equation 9, $\eta_0$ is the coupling power efficiency, which corresponds to the total transmission of the grating - collimator spectral filter, while $w_F$ is the single-mode-fiber mode field radius. According to this equation, the transmission is a Gaussian function of $\lambda$ as $\delta r$ and $\delta\theta$ are linearly proportional to $\lambda$, given that the initial error is proportional to $\lambda - \lambda_c$, as shown in Equation 6 for the narrow spectral filter BW case again. We assumed the Gaussian spectral filter that the transmission to be proportional to $\exp(-a\lambda^2)$. Then, the spectral full-width-half-maximum (FWHM) is given as $2\sqrt{\ln 2/a}$. We can find $a$ by taking the second-order derivative of the exponent $-a\lambda^2$ with $\lambda$. Therefore, the spectral FWHM width of the transmission, and thus the filter BW $\Delta\lambda$ can be determined using Equation 11.

$$\Delta\lambda = \text{spectral filter BW} = 2\sqrt{\ln 2} \Big/ \sqrt{\frac{1}{2}\frac{d^2\phi}{d\lambda^2}} \quad (11)$$

Derived from $M_{AtFiber}$, we can express the translational and angular misalignment $\delta r$ and $\delta\theta$ as functions of wavelength and substitute them in Equation 10 to calculate $\phi$. By taking necessary derivatives as shown in Equation 11, the filter BW can be determined.

To make the equation more practical, we further modify it to incorporate commonly used parameters for the filter elements. The final equation for the BW is shown as equation (12) where $D_B$ is the input beam diameter incident to the grating, $D_F$ is the fiber mode field diameter (MFD), and $\Lambda$ is the grating density (number of grooves / length). For the diffraction order, the first order $m = 1$ is used.

$$\Delta\lambda = \sqrt{\frac{\ln 2}{2}} \frac{1}{\pi\Lambda} \sqrt{\frac{(D_B^2 D_F^2 \pi^2 + 16f^2\lambda_c^2)(1 - (\Lambda\lambda_c - \sin\theta_i)^2)}{D_B^2 f^2 + D_F^2(f - L)^2}} \quad (12)$$

From Equation 12, the behavior of the $\Delta\lambda$ versus the grating incident angle has been plotted for various grating densities. The center wavelengths of 1030 nm and 1550 nm are chosen as they are common operating wavelengths of Yb-doped and Er-doped fiber lasers, respectively. Here, we targeted a 4 nm bandwidth spectral filter typically used in self-similar fiber laser mode-locking.

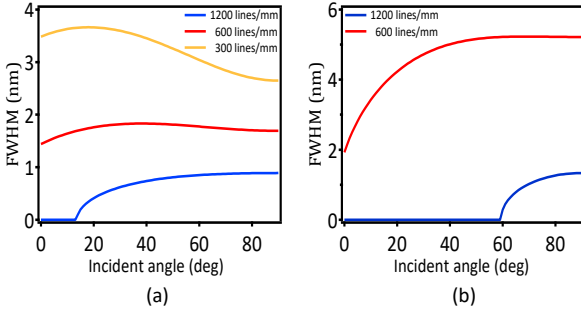

Figure 2. Filter BW vs. grating incident angle for (a) 1mm input beam diameter, 6μm fiber MFD at a wavelength of 1030nm, and (b) 0.5mm input beam diameter, 4.2μm fiber MFD at a wavelength of 1550nm.

In Figure 2(a), a typical case of a 1030 nm Yb-doped normal dispersion mode-locked fiber was studied. It was assumed that an input beam diameter of 1 mm was focused by 4.5 mm focal length lens to HI1060 fiber with ~6 μm MFD. A BW of ~3.4 nm was obtained using 300 lines/mm grating density, a 45° incident angle, and a 5 cm distance to the lens (L). This result is consistent with the previous experimental measurement of ~4 nm BW [5].

In Figure 2(b), a case for a 1550 nm wavelength Er-doped normal dispersion mode-locked fiber was examined. The calculated BW for 600 lines/mm grating at a 45° incident angle was found to be 5.1 nm with $D_B$=0.5 mm, $D_F$=6 μm, L=5 cm, and $f$=1 mm. This result agrees with the previously reported experimental measurements of a 4 nm BW [12].

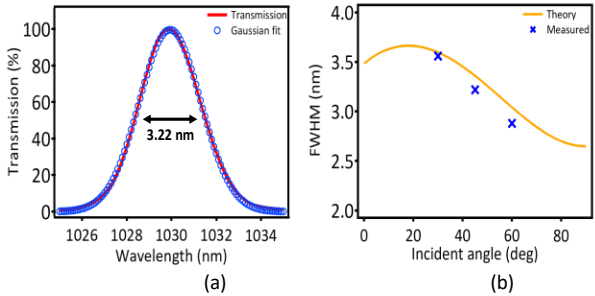

Figure 3. (a) Measured filter transmission for 45° incident angle, (b) Measured filter BW vs. theory. Conditions are the same as Figure 2(a) 300 lines/mm grating density.

We compared our calculations with calculation data through an experiment, the conditions are the same as Figure 2(a) 300 lines/mm grating density. In Figure 3(a), we measured the transmission of the filter and did the Gaussian fit to get filter BW for a 45° incident angle. By changing the incident angle to 30° and 60°, we plotted the measured FWHM value with analytical calculations as Figure 3(b). The experimental results show good agreement with the calculations.

Also, the filter BW was sensitive to variations in the input beam diameter. For example, for a 1030 nm, 1 mm beam with a 300 lines/mm grating density, a $\pm 10\%$ deviation in the beam diameter (0.1 mm) induces a $\pm 5\%$ change in BW ranging from 3.2 nm to 3.6 nm. Despite some differences between the calculated results and the previously reported experimental measurement, we anticipate that these discrepancies will decrease as the input beam diameter is carefully characterized.

## 3. Spectral filter group velocity dispersion (GVD)

The group velocity dispersion (GVD) is an important factor in the behavior of mode-locked lasers. Since the grating-collimator filter introduces GVD, it is important to assess whether the filter-induced GVD significantly impacts the mode-locked operation. GVD can be calculated from the frequency-dependent optical path length as shown in Figure 4. Since the optical pathlength changes as the grating-to-lens and lens-to-fiber distance vary, it is sensible to consider the GVD coefficient $\beta_2$ which is GVD divided by the central distance $d$, where $d = f + L$ in Figure 4. The GVD coefficient $\beta_2$ can be calculated by Equation 13, where $P$ represents the optical path length. [13].

$$\beta_2 = \frac{1}{d}\frac{\partial^2}{\partial \omega^2}\left(\frac{\omega}{c}P\right) \tag{13}$$

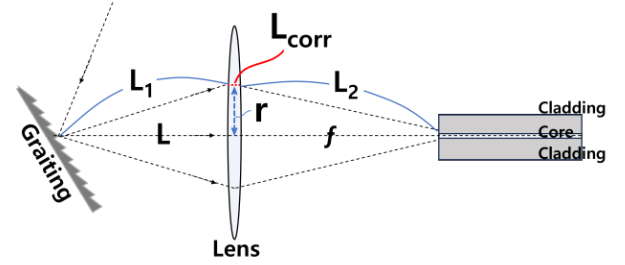

Figure 4. Group velocity dispersion calculation schematics

As illustrated in Figure 4, $P$ is the sum of the distance from the grating-to-lens $L_1$, the lens-to-fiber distance $L_2$, and a correction distance $L_{corr}$, which accounts for the index of refraction difference between free space and the lens. We assumed the lens is plano-convex.

We let $r$ and $r'$ as the beam height at the lens and fiber tip respectively. From the perspective of geometrical optics, we let L' as the distance to the image point, which is placed inside the fiber. Then, we can write $L_1 = \sqrt{L^2 + r^2}$ and $L_2 = \sqrt{L'^2 + (r - r')^2}$. To obtain the equation in terms of $f$, L, and $r$, we use a thin lens (Equation 14) and similarity relation (Equation 15).

$$\frac{1}{L'} = \frac{1}{f} - \frac{1}{L} \tag{14}$$

$$\frac{r - r'}{f} = \frac{r}{L'} \tag{15}$$

Then we get $L_2 = \sqrt{(1 - f/L)^2 r^2 + f^2}$. To find $L_{corr}$, we assumed the plano-convex lens, which has the entrance side as a plane, and the exit side as a spherical surface with a radius of curvature $R$. In this situation, spherical lens equation can be written as Equation 16.

$$\frac{1}{L'} + \frac{1}{L} = -(n-1)\frac{1}{R} \tag{16}$$

If we let $\Delta t$ to be the distance passed through the lens, $\Delta t$ is given by $R - \sqrt{R^2 - r^2} \simeq r^2/2R$ and $L_{corr} = (n-1)\Delta t$. Using the Equations 14 and 16 to replace L' and $R$ into $f$ and $r$, $L_{corr}$ can be obtained as $-r^2/2f$. Here, $r$ is the distance from the point on the lens where the central wavelength $\lambda_c$ is incident. After some mathematical manipulation, $P$ can be expressed as Equation 17.

$$P = \sqrt{L^2 + r^2} + \sqrt{\left(\frac{L-f}{L}\right)^2 r^2 + f^2} - \frac{r^2}{2f} \quad (17)$$

Since $r$ is a function of wavelength $\lambda$, by applying Equations 6, 13, and 17, $\beta_2$ can be derived, as shown in Equation 18.

$$\beta_2 = \frac{1}{f+L} \frac{\lambda^3 \sec^2(\theta_m(\lambda))(f-L)}{(2\pi c^2 \Lambda^2)} \quad (18)$$

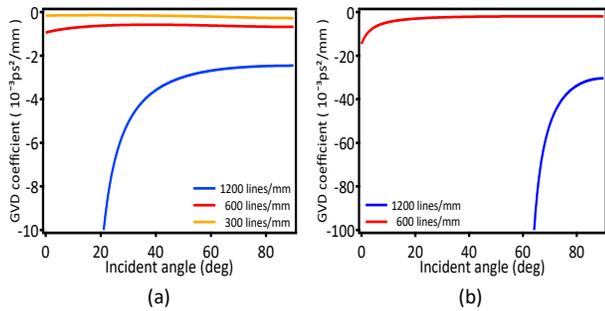

Figure 5. GVD vs. the grating incident angle. Conditions are the same as in Figure 2.

In Figure 5, the GVDs as a function of the incident angle were plotted for 1030 nm and 1550 nm center wavelengths. The negative GVD values indicate that the grating-collimator filter induces an anomalous GVD. For the 1030 nm case with a 300 lines/mm grating, the GVD effect of the filter is minimal. However, as the grating density increases, the GVD effect escalates rapidly. The study suggests that a 1200 lines/mm grating is not ideal for a normal dispersion Yb-fiber laser. Similarly, for 1550 nm wavelength, designing a normal dispersion laser with a 1200 lines/mm grating filter is impractical. This study provides insights for designing mode-locked fiber lasers.

## 4. Conclusion

Using ABCDEF matrix analysis, we analytically calculated the spectral filter BW and GVD in the diffraction grating-collimator spectral filters. The calculated BWs were compared with previously reported experiments and the results showed good agreement. For the GVD, our calculations indicate that the GVD impact is negligible at 1030 nm with a 300 lines/mm grating. Additionally, the analysis suggests that gratings with high density are impractical in normal dispersion fiber lasers.